\documentclass[aps,prl,showpacs,preprintnumbers,twocolumn,nofootinbib,superscriptaddress]{revtex4-1}

\usepackage{geometry}                % See geometry.pdf to learn the layout options. There are lots.
\usepackage{graphicx}
\usepackage{amssymb}
\usepackage{dsfont}
\usepackage{epstopdf}
\usepackage{color}
\usepackage{mathtools}
\usepackage{cancel}

\definecolor{red}{rgb}{1,0,0}
\definecolor{blue}{rgb}{0,0,1}
\definecolor{black}{rgb}{0,0,0}
\newcommand{\blue}{}

\newcommand{\p}{\partial}
\newcommand{\eq}[1]{\begin{align}#1\end{align}}
\newcommand{\eqs}[1]{\begin{align*}#1\end{align*}}
\newcommand{\ffrac}[2]{\mbox{$\frac{#1}{#2}$}}
\newcommand{\half}{\mbox{$\frac{1}{2}$}}
\newcommand{\OO}{\mathcal{O}}

\newcommand{\D}{{\mathcal{D}}}

\newcommand{\Phid}{{\Phi^\dagger}}
\newcommand{\phid}{{\phi^\dagger}}

\newcommand{\oW}{{\overline W}}

\usepackage{calrsfs}

\usepackage{scalerel,stackengine}
\stackMath
\newcommand\widecheck[1]{%
\savestack{\tmpbox}{\stretchto{%
  \scaleto{%
    \scalerel*[\widthof{\ensuremath{#1}}]{\kern-.6pt\bigwedge\kern-.6pt}%
    {\rule[-\textheight/2]{1ex}{\textheight}}%WIDTH-LIMITED BIG WEDGE
  }{\textheight}% 
}{0.5ex}}%
\stackon[1pt]{#1}{\scalebox{-1}{\tmpbox}}%
}

\newcommand{\ot}{\overline{\theta}}

\renewcommand{\oe}{\overline{\eta}}

\newcommand{\1}{\mathds{1}}

\begin{document}
\title{Fermionic theory of nonequilibrium steady states}
\author{Eric De Giuli}
\author{Masanari Shimada}
\affiliation{Department of Physics, Toronto Metropolitan University, M5B 2K3, Toronto, Canada}

\begin{abstract}
As the quantification of metabolism, nonequilibrium steady states play a central role in living matter, but are beyond the purview of equilibrium statistical mechanics. {\blue Here we develop a fermionic theory of  nonequilibrium steady states in continuous-time Markovian systems, generalizing Boltzmann-Gibbs statistical mechanics to this case. The response to an arbitrary perturbation is computed, and simplified in canonical cases. Beyond response, we consider ensembles of nonequilibrium steady states and show that a general class of ensembles is described by a 2D statistical field theory with infinitesimally broken supersymmetry, 
%and derive a fluctuation-response relation over a non-equilibrium ensemble. Some connections to quantum gravity are pointed out, and the formulation is extended to a supersymmetric integral one, 
 which may form the basis of nontrivial solvable models of nonequilibrium steady states. }

\end{abstract}
\maketitle

Many out-of-equilibrium systems can usefully be represented as Markovian, as often considered in stochastic thermodynamics \cite{Seifert12,Fang19}. %with a transition matrix $W$. %In the framework of stochastic thermodynamics, one can define, for any Markov model, entropy production, work, and energy, thus importing notions from thermodynamics to a more general setting. 
Of particular interest are non-equilibrium steady states (NESS), which do not satisfy detailed balance, and therefore cannot be studied with Boltzmann-Gibbs statistical mechanics. It is important to develop, as fully as possible, alternative methods of analysis of NESS. There exists a general diagrammatic method  \cite{Hill66,Schnakenberg76}, but it can be unwieldy; algebraic methods would be preferable. Recent work has focused on fluctuation-dissipation relations \cite{Prost09,Seifert10,Baiesi13} and response functions \cite{Owen20}, derived using idiosyncratic methods. In this letter we show how anti-commuting Grassmann variables, used in quantum field theory to represent fermions \cite{Zinn-Justin96}, can be leveraged to represent the stationary probabilities of a general NESS. To illustrate the method, we re-derive a recent and surprising result of Owen, Gingrich, and Horowitz \cite{Owen20}: if a multiplicative perturbation is made to all transition rates out of state $j$, then the response of all stationary probabilities $\pi_i$ takes an explicit universal form, depending only on the stationary probabilities before the perturbation. We then extend their result to give the response to arbitrary observables (see also \cite{Aslyamov23} ). Going beyond response, we build a supersymmetric representation of NESS, which allows analysis of ensembles of NESS. {\blue We show that a general class of ensembles with both topological and rate disorder are described by 2D statistical field theories with infinitesimally broken supersymmetry (SUSY), adding to evidence that nontrivial phases of disordered systems are described by SUSY-breaking theories with instanton excitations. }\\

%We consider, under some approximations, ensembles over random forces, and derive a fluctuation-response relation over non-equilibrium ensembles. A mapping to 2D quantum gravity is made. \\

%At a crude level a NESS is characterized by the probability $\pi_i$ to be in state $i$. As a first step in understanding a NESS, one would like to know its response to various perturbations. Recently, XXXX derived a surprising result: if a perturbation is made to all transition rates out of state $j$, then the response of all stationary probabilities $\pi_i$ takes an explicit universal form, depending only on the stationary probabilities. The proof in XXX was obtained by explicit counting of graphs in state space, utilizing Kirchoff's matrix-tree theorem \cite{Kirchhoff47}. In this short note we show how the same result can be obtained with the Berezin integral, which is used in quantum field theory to represent fermionic degrees of freedom.

{\bf Markov models and the matrix-tree theorem: }
Consider a general continuous-time Markovian system with a transition matrix $W_{ij}$ for the rate that $j \to i$. Define
\eq{
R_{ij} = \begin{cases} W_{ij} & i \neq j \\ -\sum_{k\neq i} W_{ki} & i=j \end{cases}
}
Then the dynamics is governed by the master equation
%\eq{
$\p_t p_i = R_{ij} p_j, $
%}
which preserves probability since $\sum_i R_{ij} = 0$; % R_{jj} + \sum_{i \neq j} R_{ij} = -\sum_{k\neq j} W_{kj} +\sum_{i\neq j} W_{ij} = 0$. 
 We make two assumptions on the rates: (A1) the process is reversible: if $W_{ij} \neq 0$, then $W_{ji} \neq 0$; and (A2) that every pair of states $i$ and $j$ are connected by a path $i \to k_1 \to k_2 \to \ldots \to k_\ell \to j$ for some $\ell$. If (A2) didn't hold in our original system, the state space could be decomposed into independent subsets. This assumption requires that we have identified all conserved quantities and conditioned the state space on their values.  

Assuming (A1,A2) there exists a unique stationary state $\pi_i$ (Kirchhoff \cite{Kirchhoff47}) \cite{Schnakenberg76,Van-Kampen92}, whose form is explicitly found using the matrix-tree theorem \cite{Schnakenberg76,Chaiken82,Caracciolo04}. We consider the state-space graph $G$, which is connected by (A2). Note that there are no edges from $i$ to $i$. Let there be $N$ vertices. A spanning tree $T$ of the graph $G$ is a tree with $N$ vertices. It therefore passes through all vertices and has $N-1$ edges. For any spanning tree $T$ and vertex $i$ we define $T_i$ as the tree with all edges directed towards $i$. This notion is well-defined since $T$ is a tree. Label these spanning trees by $T^\mu_i$. Let $A(T^\mu_i) = \prod_{e \in T^\mu_i} W_e$, where $e$ labels directed edges, e.g if $e = (j \to k)$ then $W_e = W_{kj}$. Kirchhoff's result is that
\eq{ \label{pi1}
\pi_i = S_i/S,
} 
where the state amplitude
\eq{
S_i = \sum_\mu A(T^\mu_i)
}
is the sum over all spanning trees directed towards $i$, and $S = \sum_i S_i$. From this representation we immediately see that the solution is guaranteed to have all positive elements, as required physically, and is normalized.

Write $-R^{(i_1 i_2 \cdots | j_1 j_2 \cdots)}$ for the matrix where rows $i_1,i_2,\ldots$ and columns $j_1,j_2,\ldots$ are removed from $-R$. {\blue The cofactor
\eq{
(i_1 \cdots i_\ell | j_1 \cdots j_\ell) \equiv (-1)^{\sum_p (i_p+j_p)} \det (-R^{(i_1 i_2 \cdots | j_1 j_2 \cdots)})
}
will play a central role in what follows. It has a diagrammatic interpretation (All-minors matrix-tree theorem \cite{Chaiken82}) \footnote{Chaiken \cite{Chaiken82} defines his $A$ matrix as our $-R$ matrix, but uses $A_{ij}$ for the arc $i\to j$, which is opposite to our convention. Thus we need to flip the signs on the arcs in his results.} as a sum over forests of $\ell$ directed trees $T_a$, each of which contains one of the $i$ vertices and one of the $j$ vertices, with all edges pointing towards the $j$ vertices, a weight $A(T_a)$, and a sign $\epsilon(\pi^*)$ that will not be used below. Note that within the lists of rows or columns, the ordering of indices doesn't matter. By convention, when the same index appears twice in the list of rows (or columns), we define the result to be zero. \\

Kirchhoff also proved that
\eq{
S_i = (i|i) = \det (-R^{(i|i)}),
}
which is used below. Eq.\eqref{pi1} generalizes Boltzmann-Gibbs statistical mechanics to nonequilibrium steady states. The tree sums $S_i$ play the role of Boltzmann weights, and $S$ plays the role of the partition function. Out of equilibrium, the probability of a state depends on the entire suite of transition rates, allowing richer behavior than in the equilibrium case. Below we will see how this leads to new dynamical DOF to describe NESS.
}

{\bf Grassmann-Berezin integrals: } Kirchoff's result relates to fermions in that determinants have a natural representation in terms of anticommuting Grassmann variables, satisfying
%\eq{ \label{anti}
$\theta_i \theta_j = - \theta_j \theta_i $.
%}
In particular, $\theta_i^2=0$. Although anticommuting variables have explicit representations as matrices, it is useful to consider them as a formal algebraic objects, subject to Eq.\eqref{anti}; for introductions see \cite{Itzykson91,Zinn-Justin96}. We follow the notation and definitions from \cite{Zinn-Justin96}.

Berezin integrals are defined by $\int d\theta_i 1 = 0$ and $\int d\theta_i \theta_i = 1$. For a function $f(\theta)$ with Taylor expansion $f(x)=f_0+ x f_1 + \half x^2 f_2 + \ldots$ we have $f(\theta_i)=f_0+ \theta_i f_1$, using $\theta_i^2=0$. Its integral is $\int d\theta_i f(\theta_i) = f_1 = f'(0)$. Integration is thus equivalent to differentiation for Grassmann variables.

We introduce a second independent set of variables $\ot_i$, anticommuting with the $\theta_i$. Define the partition function with anti-commuting fields $\eta, \oe$ as 
\eq{ \label{Z1}
%Z[\eta,\oe] = \int \D\theta\D\ot e^{\sum_{i,j} M_{ij} \ot_i \theta_j + \oe_i \theta_i + \ot_i \eta_i} , 
Z[\eta,\oe] = \int \D\theta\D\ot \; e^{\ot \cdot M \cdot \theta + \oe \cdot \theta + \ot \cdot \eta} , 
}
where $\D\theta\D\ot \equiv \prod_i d\theta_i d\ot_i$. For arbitrary $M$ we have $Z[0,0] = \det M$, and for non-singular $M$ we have $Z[\eta,\oe]= \det M e^{-\oe_i M^{-1}_{ij} \eta_j}$. We define unnormalized expectation values 
\eq{
[ \ot_{i_1} \theta_{j_1} \cdots \ot_{i_n} \theta_{j_n}  ]_M \equiv \int \D\theta\D\ot e^{\ot \cdot M \cdot \theta } \ot_{i_1} \theta_{j_1} \cdots \ot_{i_n} \theta_{j_n} 
}
with the convention that if no matrix $M$ is indicated, then $M=-R$. From the above it follows
\eq{ \label{pair}
[\ot_i \theta_j]_M = (\det M) M^{-1}_{ji}
}
when $M$ is invertible. The Grassmannian Wick theorem is \cite{Zinn-Justin96}
\eqs{
& (\det M)^{n-1} [ \ot_{i_1} \theta_{j_1} \cdots \ot_{i_n} \theta_{j_n}  ]_M \\
& \qquad = \sum_P \epsilon(P) [\ot_{i_1}\theta_{j_{P_1}}]_M \cdots [\ot_{i_n}\theta_{j_{P_n}}]_M ,
}
where the sum is over all permutations $P$ of $\{ j_1,j_2,\ldots,j_n \}$, with $\epsilon(P)$ its sign. \\

{\bf Determinant \& sum identities: }
Since expectation values are polynomials in the entries of $M$, they can be computed for arbitrary matrices, singular or not. Setting $M=-R$ the left-hand-side of the Wick theorem vanishes and we see immediately that the quantities $[\ot_i \theta_j]$ are subject to numerous identities. Moreover, by applying \eqref{Z1} with $M=-R$ (at zero field) we get a nontrivial representation of $0=\det(-R)$. Using $\p R_{pq}/\p W_{ij}= \delta_{pi}\delta_{qj}-\delta_{pq}\delta_{qj}$ and differentiating $\det(-R)$ with respect to a rate we find
\eq{
0 = \frac{\p \det(-R)}{\p W_{ij}} = [\ot_i \theta_j] - [\ot_j \theta_j] ,
}
reducing all such expectations to the `diagonal' ones. 

%We can go further: 
%\eq{
%0 = \frac{\p^2 \det(-R)}{\p W_{kl} \p W_{ij}} = [(\ot_i-\ot_j) \theta_j (\ot_k-\ot_l) \theta_l] 
%}
%and so on. 

To actually evaluate expectations, we cannot apply the Wick theorem with $-R$, since the left-hand side vanishes. But we can apply the theorem with $M=-R + t I^q$, where $(I^q)_{jk} = \delta_{jk}\delta_{qk}$, and then set $t \to 0$ later. By Laplace expansion we get (see SI)
\eqs{
[\ot_i \theta_j]_M & = (i | j) + t (iq | jq)
}
which leads to
\eq{
[\ot_i \theta_j \ot_k \theta_l ] & = \frac{(i|j)(kq|lq)+(k|l)(iq|jq)-(i|l)(kq|jq)-(k|j)(iq|lq)}{(q|q)} ,
}
for any choice of the index $q$. These relations hold also if some indices are equal, provided minors with repeated indices are set to 0, which is the Pauli exclusion principle. From this we can derive numerous identities. Below we will use 
\eq{ \label{Q}
[\ot_i \theta_i (\ot_k-\ot_l) \theta_l ] = \frac{1}{2(k|k)} \big[ Q^i_{kl} - (i|i) (kl|kl) \big] ,
}
where $Q^i_{kl}=(l|l)(kl|ik)-(k|k)(kl|il)$ is odd in its lower indices: $Q^i_{kl}=-Q^i_{lk}$. This is useful when $k$ and $l$ come from an edge connecting two states, and $i$ is a `target' state, because $Q$ has a definite parity with respect to edges, and the dependence on $i$ is disentangled as much as possible. 

We can similarly derive a useful sum identity
\eq{ \label{sum1}
\sum_i (iq|jq) R_{ik} = (j|j) \delta_{kq} - (q|q) \delta_{jk}
}
{\blue By the all-minors matrix-tree theorem, all these identities have nontrivial diagrammatic interpretations. Yet they effortlessly follow from the fermionic approach, and infinitely more can be derived at higher orders.  }

{\bf Response: } Kirchoff's result reads $S_j = (j|j)$. We thus have the fermionic representation
\eq{
S_j = (j|j) = [\ot_j \theta_j] = [\ot_i \theta_j] ,
}
for any $i$. We now use this to compute response functions. Following \cite{Owen20} we parametrize rates as
\eq{
W_{ij} = e^{E_j-B_{ij}+F_{ij}/2} ,
}
where $B_{ij}=B_{ji}$ and $F_{ij}=-F_{ji}$. In this Arrhenius-like form, the $B$ are interpreted as dimensionless barrier heights \cite{Mandal11}, while the $F$ correspond to dimensionless forces \cite{Aslyamov23}. If the rates depend on $\lambda$ then in general we have
\eq{ \label{resp1}
\frac{\p \log \pi_i}{\p \lambda} = \frac{\p \log S_i}{\p \lambda} - \frac{\p \log S}{\p \lambda} ,
}
where
\eq{ \label{resp2}
\frac{\p \log S_i}{\p \lambda} & = \frac{1}{S_i} \frac{\p [\ot_i \theta_i]}{\p \lambda} = -\frac{1}{S_i} \sum_{k,l} [\ot_i \theta_i \ot_k \theta_l] \frac{\p R_{kl}}{\p \lambda} \notag \\
& = -\frac{1}{S_i} \sum_{k,l} [\ot_i \theta_i (\ot_k-\ot_l) \theta_l] \frac{\p W_{kl}}{\p \lambda} ,
}
{\blue which can be used to compute arbitrary response functions. Using this relation and the above identities, we can derive the responses to vertex \cite{Owen20}, barrier \cite{Aslyamov23}, and force \cite{Aslyamov23} perturbations. They read (see SI)
\eq{
\frac{\p \log \pi_i}{\p E_p} & = \pi_p - \delta_{ip} \\
\frac{\p \log \pi_i}{\p B_{kl}} & =  \left( -\frac{Q^i_{kl}}{2 S_i S_k S_l} + \sum_q \frac{Q^q_{kl}}{2 S_q S_k S_l} \right) J_{kl}  \\
\frac{\p \log \pi_i}{\p F_{kl}} & = \left( \frac{Q^i_{kl}}{4 S_i S_k S_l} - \sum_q \frac{Q^q_{kl}}{4 S_q S_k S_l} \right) \tau_{kl} ,
}
in terms of the flux $J_{kl} = W_{kl} \pi_l - W_{lk} \pi_k$ from $l$ to $k$ and the edge traffic \cite{Maes20} $\tau_{kl} = W_{kl} \pi_l + W_{lk} \pi_k$ between states $l$ and $k$.

}

Similar algebraic expressions for $\frac{\p \log \pi_i}{\p B_{kl}}$ and $\frac{\p \log \pi_i}{\p F_{kl}}$ were used in \cite{Aslyamov23} to obtain bounds on the sensitivities of edge currents and traffic.

Further response functions are obtained straightforwardly from \eqref{resp1} \& \eqref{resp2}, and simplified with the determinant and sum identities as needed. With them, one can compute the entropy difference in slow transitions between nonequilibrium states \cite{Mandal16}.\\

{\bf Nonequilibrium Ensembles: }

%{\blue
%$(S_i)^{-1} = t (\det (-R + t I^i) )^{-1} = t \int \D s\D u \; e^{-R : (i s u)} e^{i t s_i u_i}$ 
%}

Since the fermionic representation is also an integral representation, it is convenient for performing ensemble calculations, {\blue in which the transition rates are themselves considered random variables. We consider both topological disorder, whereby the states are connected in an inhomogeneous way, and disorder in the values of the rates. Ensembles are commonly used to understand the general phase behavior of complex systems \cite{Mezard87,Parisi99,DeGiuli19,Mosam21,De-Giuli22b}, since many observables become independent of the disorder realization in the large system limit.  %The latter can have several interpretations: most simply, we can consider ensemble averages over some parameters under experimental control, or subject to external fluctuations; alternatively, we may be interested in a single large system, and hope to extract universal behavior by taking a thermodynamic limit. 
%The latter approach is 
% Here we illustrate the latter approach, commonly used to understand the general phase behavior of complex systems . 
 We will show that the NESS of a very large class of systems with topological and/or transition-rate disorder are described by a 2D statistical field theory with infinitesimally broken supersymmetry. 

%Here we illustrate both uses. For simplicity, we restrict ourselves to $\langle S_i \rangle$ over ensembles; later we will discuss extensions to $\langle \pi_i \rangle$. First, consider a randomly driven system with Gaussian $F_{kl}$, $\langle F_{kl} \rangle=0, \langle F_{kl}^2 \rangle = \delta F_{kl}^2$, iid for each $F_{kl}$. It follows by a simple computation that $\langle S_i \rangle$ is exactly described by an effective {\it detailed-balanced} system with renormalized barriers $B_{kl} \to B_{kl} - \ffrac{1}{8} \delta F_{kl}^2$. However, this simple result (which is only approximate, see below) only applies to single-state statistics. If we compute the correlations $\langle S_i S_j \rangle_c=\langle S_i S_j \rangle-\langle S_i\rangle\langle S_j \rangle$ we find {\blue to leading order in the forces }
%\eq{
%\langle S_i S_j \rangle_c = 4 \sum_{k<l} (e^{\frac{1}{4}\delta F_{kl}^2}-1) \frac{\p S_i}{\p F_{kl}} \frac{\p S_j}{\p F_{kl}} ,
%}
%which is a type of fluctuation-response relation, over this ensemble of NESS. We emphasize that the ensemble is not explicitly dynamical: how it is explored is immaterial. Similar expressions will hold when arbitrary control parameters are given Gaussian fluctuations.

We define a {\it reactive ensemble} as follows. Let there be $N$ states, of $K$ different types labeled by $\mu$. Transition rates depend only on the type of state; for each pair $\mu,\nu$ we have transition rates $W_{\mu\nu},\oW_{\mu\nu}$ for the transition rates from $\nu\to \mu$ and $\mu \to \nu$, respectively. When $\mu \neq \nu$ we assume that these transitions pass through `activated' states. We can arbitrarily specify the number of each type of state $N_\mu$, the number of transition channels to other states, $E_\mu$, (which may include transitions to other states of the same type, and self-transitions), and the number of transitions through activated states, $E_{\mu\nu}$. Then the reactive ensemble is the ensemble over all possible ways to connect the states, subject to these conditions. Moreover, the rates $W_{\mu\nu},\oW_{\mu\nu}$ can themselves be random and their effects averaged over in the ensemble.

%Using the fermionic formalism, we can show that any such reactive ensemble is described by a 2D statistical field theory with infinitesimally broken supersymmetry. 

%The superfield $\Phi^\mu$ has $K$ components. Moreover, when $N=\sum_\mu N_\mu \gg 1$ and $K$ is finite, the theory is governed by saddle points. 

A few examples are as follows: if $K=1$ then all states are equivalent after averaging and we have a mean-field nonequilibrium disordered system; in \cite{Mosam21} it was shown that when the rates have a wide distribution, many features of human fMRI data are well-described by such an ensemble, without fitting parameters. If $K>1$ then we distinguish types of states. These could be `excited' versus `quiescent', `supercritical' versus `subcritical', and so on, possibly with further grading. If $K \gg 1$ then we can specify occupation numbers of `molecules', etc. 

In the main text, we present results for $K=1$ and topological disorder only; the straightforward extension to $K>1$ and with rate disorder is shown in Supplementary Material, along with the proofs.

Consider a general observable of rate pairs $\pi_i \OO(W_{ij},W_{ji})$. In the NESS we want this weighted with the steady-state probabilities, averaged over the ensemble: $\langle \pi_i \OO(W_{ij},W_{ji}) \rangle$. For example, in stochastic thermodynamics $\pi \OO$ could be the entropy production in the reservoir $\pi_i W_{ji}\log W_{ji}/W_{ij}$, or moments of the rates, measuring clustering in the state space, etc. 

The first step is to check that the matrix $M' = -R+t \1$, with $\1{}_{jk}=1$ for all $j,k$, has $\det(M')=t N S$; then we use a bosonic representation of $1/|\det M'|$. Since $\det M'=t NS>0$ (for $t>0$), we avoid the sign problem that plagues other approaches to disordered systems \cite{Parisi79,Kurchan02,Tissier11,Tarjus13,Tarjus20}. This leads to
\eq{
\frac{\pi_ i}{N}  = \int \D\theta\D\ot \int \D s\D u \; e^{-R : (\ot \theta + i s u)} e^{t \ot_i \theta_i + i t s \cdot \1 \cdot u} ,
}
for any positive\footnote{Negative values of $t$ can also be used, if $\pi_i$ is multiplied by an additional factor sgn$(t)$.} value of $t$, where $\D s\D u = (2\pi)^{-N} \prod_j ds_j du_j$. It can easily be verified that at $t=0$ the action is invariant under the BRST transformation \cite{Zinn-Justin96} $\delta s_j = \epsilon' \ot_j, \delta u_j = \epsilon \theta_j,  \delta \ot_j = -i \epsilon s_j, \delta \theta_j = i \epsilon' u_j$, where $\epsilon$ and $\epsilon'$ are independent Grassmann numbers. Thus at $t=0$ the action is supersymmetric; but, crucially, the SUSY is broken by the `fields' proportional to $t$; indeed, without these terms, we would have a representation of the ill-defined quantity $\det(-R)/\det(-R)=0/0$.

Then, the next step is to represent the right-hand side, multiplied by $\OO$, as the amplitude of a Feynman diagram \cite{Bachas94,DeGiuli19a}. States will be vertices of the diagram; then to each state is associated two Grassmann dimensions $\theta_j,\ot_j$ and two real dimensions $s_j,u_j$ (whose product has units of time). After disentangling $e^{i t s \cdot \1 \cdot u}  = \int \frac{dxdy}{\pi} e^{-(x^2+y^2)} \prod_k e^{i \sqrt{t} (x+iy) s_k + \sqrt{t} (x-iy) u_k } \equiv \int \frac{dxdy}{\pi} e^{-(x^2+y^2)} \prod_k \lambda(s_k, u_k)$ we need to give states a contribution $\lambda(s_k,u_k)$ and edge pairs their contribution to $e^{-R : (\ot \theta + i s u)}$. In terms of the complex scalar superfield $\Phi$ 
%
%In terms of the complex scalar superfield 
%\eq{
%\Phi(\ot,\theta,s,u) = \phi(s,u) + \ot \epsilon(s,u) + \bar\epsilon(s,u) \theta + \ot \theta \psi(s,u)
%}
and supercoordinate $R = (\ot,\theta,s,u), T = (\oe,\eta,v,w)$ this is accomplished by the action
\eq{ \label{S}
S(\Phi) & = \int dR \int dT \Phid(R) M(R,T)\Phi(T) \notag\\
& - \int dR \sum_{n \geq 1} \frac{1}{n!} \xi_{n} \lambda(R)  (c\Phi(R)+c^\dagger \Phid(R))^n 
}
where 
\eqs{
M(R,T) = \frac{\delta(v-s)\delta(w- u \ffrac{W}{\oW})}{-\oW/(2\pi)} e^{(\oe-\ot) (W\theta-\oW \eta)-i \p_{v} \p_{w} / \oW} .
}
Briefly, the Wick expansion with respect to $\xi_n$ creates vertices (states) with $n$ neighbors, and the inverse propagator $M$ is designed to give the correct edge-pair contribution when inverted. Contour integrals over $\xi_n$ allow one to exactly control the degree distribution. When $N\to \infty$, observables are extracted by evaluating on saddle-points, leading to %$\langle \pi \OO  \rangle \sim N\int \frac{dxdy}{\pi } e^{-x^2-y^2} \tilde \OO[\Phi]$ where $\tilde\OO[\Phi]$ is the translation of the observable into the field theory (see SI), and $\sim$ means `asymptotic to'.
$\langle \pi \OO \rangle \sim \int \frac{dxdy}{\pi/N} e^{-x^2-y^2} \left(\mathcal{W}/J\right)_{J=0}$, where $\mathcal{W}$ is an appropriate generating function (see SI), and $\sim$ means `asymptotic to'.

Eq.\eqref{S} has several fundamental properties. First, when $t=0$ it is supersymmetric, and since the amplitude of $t$ is arbitrary, it can be made infinitesimal. There will be phases in which the limit $t\to 0$ is benign and SUSY is preserved; then the $x,y$ integrals are trivial. Elsewhere, SUSY will be broken and $\Phi$ will pick up explicit $x,y$ dependence. Second, when $W\neq \oW$, detailed balance is broken and the theory is {\it nonlocal}. Moreover, the kinetic operator $e^{-i \p_{v} \p_{w} / \oW}$ involves derivatives of all orders. 

These results may shed light on a longstanding debate in the theory of disordered systems. Note first that the origin of SUSY here is that we are representing $\pi_i=S_i/S$ as a ratio whose numerator becomes fermionic and whose denominator becomes bosonic: so any change in the former must be compensated by a change in the latter, to preserve conservation of probability. The origin of the two real dimensions is that they capture fluctuations in $1/S$, which are generally significant since $S$ is exponentially large in the number of states. If we had replaced $1/S$ by $1/\langle S \rangle$, then these fluctuations would be neglected and the theory would become simple, in analogy to the annealed approximation \cite{Mezard87}, highlighting again the analogy between $S$ and the partition function of equilibrium systems. Since spatial structure is lacking, these real dimensions must be temporal; we expect that $\Phi$ contains information on correlation and response functions \cite{Cugliandolo99}.

The low-temperature marginal phase of mean-field spin glasses (and other disordered systems \cite{Biroli18,Parisi20} ), where the annealed approximation breaks down, is described by replica symmetry breaking \cite{Parisi80,Parisi80a,Parisi80b,Mezard87}. It has long been an open question what is the correct field-theoretic description of the marginal phase, a prerequisite to understanding its behavior under renormalization. Replica symmetry is equivalent to SUSY in dynamic treatments \cite{Kurchan92}, and breaking of SUSY has been linked to breaking of time-reversal invariance \cite{Sompolinsky82,Kurchan91,Kurchan92}, but explicit computations in the SUSY-broken phase have remained limited \cite{Bray80,Parisi04}.

The marginal phase is critical: avalanches have scale-free distributions \cite{Pazmandi99,Doussal10,Le-Doussal12,Muller14}, as expected to be ubiquitous in disordered systems \cite{Bak87,Muller14}. It was proposed in \cite{Ovchinnikov11,Ovchinnikov11a,Ovchinnikov16} that self-organized criticality corresponds to a Witten-type topological field theory \cite{Witten81} with spontaneously broken SUSY. Several features of the present theory agree with nontrivial predictions of \cite{Ovchinnikov11,Ovchinnikov11a,Ovchinnikov16}: first, since Eq.\eqref{S} is dominated by saddle-points, fluctuations can only be relevant via instantons, or defects, as proposed in \cite{Ovchinnikov11}. Second, here we can see explicitly the form of the SUSY-breaking terms, in $\lambda(R)$; they break the homogeneity of the real dimensions, as also proposed in \cite{Ovchinnikov11}. Thus Eq.\eqref{S} looks promising to obtain exactly solvable models of NESS as dynamically broken SUSY. 

Moreover, this theory can be further extended. If $\Phi$ is promoted to a matrix valued field, then the state space becomes locally two dimensional. A version of this theory, in the annealed approximation and with uniform transition rates, was argued to be a discretization of 2D quantum gravity with fermionic matter, with partition function $Z_{QG}$ \cite{Kazakov85,David85,Gorsky23}. Comparing to the model of \cite{Gorsky23}, one can easily prove that
\eq{ \label{QG}
\langle\left. S \right|_{W=A=A^T}\rangle  = -\left.\frac{\p Z_{QG}}{\p m^2} \right|_{m^2=0} 
}
where the average is over all undirected planar graphs with adjacency matrix $A$, and where $m$ is the mass of the fermions; thus the tree partition function $S$ is that of 2D quantum gravity with massless spinless fermions\footnote{or pure gravity, depending on the way the limit is taken, and up to an overall irrelevant multiplicative constant.}. This relationship is known and used in the solution of the model, but the precise connection to $\pi_i=S_i/S$ indicates its connection to nonequilibrium physics. $Z_{QG}$ can be evaluated by mapping to a Hermitian matrix model \cite{Di-Francesco95,Eynard16}, whose Feynman diagrams reproduce  Kirchoff's spanning trees.

}

{\bf Conclusion: } {\blue We present a fermionic and eventually supersymmetric representation of nonequilibrium steady states, generalizing Boltzmann-Gibbs statistical mechanics. }With the aid of various identities following from the singular nature of the transition rate matrix, the general response function is derived in \eqref{resp1} \& \eqref{resp2} (see also \cite{Aslyamov23}), agreeing with previous work on vertex \cite{Owen20} and edge \cite{Aslyamov23} perturbations. The representation is extended to a supersymmetric integral one, suitable for analysis of ensembles. {\blue A large class of ensembles is exactly described by a 2D statistical field theory with infinitesimally broken supersymmetry, while an ensemble over planar state spaces maps onto 2D quantum gravity. 
%An ensemble over random forces shows a fluctuation-response relation, while an ensemble over planar state spaces maps onto 2D quantum gravity. 
In future work it would be informative to find solvable models of genuine nonequilibrium ensembles, to show in detail how replica symmetry breaking corresponds to SUSY-breaking.}

{\bf Acknowledgments: } We are grateful to N. Freitas for comments on an early version of the manuscript. EDG is supported by NSERC Discovery Grant RGPIN-2020-04762. 

%\bibliographystyle{abbrv}
%\bibliography{../Gravity,../Glasses}
\bibliography{../../language,../../Biology,../../Glasses}

\vfill

\begin{widetext}

{\bf Supplementary Material. } \\

{\bf 1. Fermionic identities. }

We have, by Laplace expansion,
\eq{
\det(-R + t I^q) & = t \det(-R^{(q|q)}) + \det(-R) \notag \\
& = t (q|q) ,
}
and, using $\theta_j^2=0$ 
\eqs{
[\ot_i \theta_j]_M & = \ffrac{1}{t} [e^{t \ot_i \theta_j}-1]_M \\
& = \ffrac{1}{t} \left( \det (M + t \vec{e}^i \vec{e}^j) - \det M \right) \\
& = \ffrac{1}{t} \left( \det M + t (-1)^{i+j} \det (M^{(i|j)}) - \det M \right) \\
& = (-1)^{i+j} ( \det (-R^{(i|j)}) + t \det (-R^{(iq|jq)}) ) \\
& = (i | j) + t (iq | jq)
}
where $\vec{e}^i$ is the vector $e^i_k = \delta_{ik}$. Note in particular that $[\ot_i \theta_j] = (i|j) = (j|j)$ using the above identity. Thus
\eq{
& [\ot_i \theta_j \ot_k \theta_l ] = \lim_{t \to 0} \frac{[\ot_i \theta_j]_M [\ot_k \theta_l ]_M - [\ot_i \theta_l]_M[ \ot_k \theta_j ]_M}{t (q|q)} \\ 
& = \frac{(i|j)(kq|lq)+(k|l)(iq|jq)-(i|l)(kq|jq)-(k|j)(iq|lq)}{(q|q)} ,
}
for any choice of the index $q$. These relations hold also if some indices are equal, provided minors with repeated indices are set to 0, which is the Pauli exclusion principle. 

Taking successively $q=i, j, k, l$ we get 4 representations
\eq{
[\ot_i \theta_j \ot_k \theta_l ] & = \frac{(i|j)}{(i|i)}(ik|il) - \frac{(i|l)}{(i|i)} (ik|ij) \\
%& = \frac{(i|j)}{(j|j)}(jk|jl) - \frac{(k|j)}{(j|j)} (ij|jl) 
& = (jk|jl) - (ij|jl) \\
%}
%It is particularly useful to take $q=k$ and $q=l$ to give
%\eq{
%[\ot_i \theta_j \ot_k \theta_l ] 
& = \frac{(k|l)}{(k|k)}(ik|jk) - \frac{(k|j)}{(k|k)} (ik|kl) \\
%& = \frac{(k|l)}{(l|l)} (il|jl) - \frac{(i|l)}{(l|l)} (kl|jl) 
& = (il|jl) - (kl|jl) 
}
Below we will need in particular
\eq{
[\ot_i \theta_i \ot_k \theta_l ] & = (ik|il) \\
%}
%\eq{
%[\ot_i \theta_i \ot_k \theta_l ] 
& = \frac{(l|l)}{(k|k)}(ik|ik) - \frac{(i|i)}{(k|k)} (ik|kl) \\
& = (il|il) - (kl|il) 
}
which imply
\eq{
[\ot_i \theta_i (\ot_k-\ot_l) \theta_l ] = - (kl|il) .
}
Below we will use this where $k$ and $l$ come from an edge connecting two states, and $i$ is a `target' state. It is then useful to rewrite this in terms of parts with definite parity with respect to edges, and also to disentangle the dependence on $i$ as much as possible. 

Permuting indices $i \leftrightarrow l$ above we can derive an identity
\eq{
(i|i)(kl|kl)-(l|l)(kl|ik)=(k|k)(kl|il)
}
so that
\eq{
& [\ot_i \theta_i (\ot_k-\ot_l) \theta_l ] = - \gamma (kl|il) \notag\\
& \qquad - \frac{(1-\gamma)}{(k|k)} \big[ (i|i)(kl|kl)-(l|l)(kl|ik) \big] ,
}
for arbitrary $\gamma$. We see that when $\gamma=1/2$ we have
\eq{ \label{Q}
[\ot_i \theta_i (\ot_k-\ot_l) \theta_l ] = \frac{1}{2(k|k)} \big[ Q^i_{kl} - (i|i) (kl|kl) \big] ,
}
where $Q^i_{kl}=(l|l)(kl|ik)-(k|k)(kl|il)$ is odd in its lower indices: $Q^i_{kl}=-Q^i_{lk}$.

We can similarly derive a sum identity by taking $M$ as here and multiplying \eqref{pair} by $M$, to get:\eqs{
t(q|q)\delta_{jk} & = (\det M) \delta_{jk} = \sum_i [\ot_i \theta_j]_M M_{ik} \notag \\
& = \sum_i \big( i|j) + t (iq|jq) \big) M_{ik} \notag \\
& = t (j|j) \delta_{kq} - \underbrace{\sum_i (i|j) R_{ik}}_{=0} - t \sum_i (iq | jq) R_{ik} ,
}
giving
\eq{ \label{sum1}
\sum_i (iq|jq) R_{ik} = (j|j) \delta_{kq} - (q|q) \delta_{jk}
}
By the all-minors matrix-tree theorem, all these identities have nontrivial diagrammatic interpretations. Yet they effortlessly follow from the fermionic approach, and infinitely more can be derived at higher orders.  \\
%These results will suffice for us. \\%, although computations can clearly be extended to higher orders without difficulty.\\

{\bf 2. Response functions. }

The response to a vertex perturbation is
\eqs{
\frac{\p \log S_i}{\p E_p} & = -\frac{1}{S_i} \sum_{k} [\ot_i \theta_i (\ot_k-\ot_p)\theta_p] W_{kp} \\
& = -\frac{1}{S_i} \sum_{k} (ik | ip ) W_{kp} \\
& = -\frac{1}{S_i} \big( S_p \delta_{ip} - S_i \big)
}
where we used the sum identity with $(q,i,k,j) \to (i,k,p,p)$. Then
\eq{
\frac{\p \log \pi_i}{\p E_p} & = (1-\delta_{ip}) + \frac{1}{S} \sum_j \big( S_p \delta_{jp} - S_j \big) \notag\\
& = \pi_p - \delta_{ip} ,
}
which is the result of \cite{Owen20}. 

The result to a barrier perturbation is, using \eqref{Q}, 
\eqs{
\frac{\p \log S_i}{\p B_{kl}} & = -\frac{1}{S_i}  [\ot_i \theta_i (\ot_k-\ot_l) (\theta_l W_{kl} - \theta_k W_{lk}) ] \notag \\
%&  = -\frac{1}{2S_i} \left( \frac{Q^i_{kl}-S_i (kl|kl)}{S_k} \right) W_{kl} -\frac{1}{2S_i} \left( \frac{Q^i_{lk}-S_i (kl|kl)}{S_l} \right) W_{lk} \notag \\
&  = -\frac{Q^i_{kl}-S_i (kl|kl)}{2S_i S_k} W_{kl} -\frac{Q^i_{lk}-S_i (kl|kl)}{2S_i S_l}  W_{lk} \notag \\
&  = -\frac{1}{2S_i} \frac{Q^i_{kl}}{S_k S_l} \tilde J_{kl}  + C_{kl}
}
for some quantity $C_{kl}$ that is independent of $i$, where $\tilde J_{kl} = W_{kl} S_l - W_{lk} S_k$ is the unnormalized flux from $l$ to $k$. Thus
\eq{
\frac{\p \log \pi_i}{\p B_{kl}} & = \left( -\frac{Q^i_{kl}}{2 S_i S_k S_l} + \sum_q \frac{Q^q_{kl}}{2 S_q S_k S_l} \right) J_{kl} 
}
where $J_{kl}=\tilde J_{kl}/S$ is the normalized flux. As emphasized by \cite{Aslyamov23}, the response vanishes if the flux vanishes on the edge, even if the system is not globally detailed balanced. 

The response to a force perturbation is obtained in the same way:
\eq{
\frac{\p \log S_i}{\p F_{kl}} &  = \frac{1}{4S_i} \left( \frac{Q^i_{kl}-S_i (kl|kl)}{S_k} \right) W_{kl} \notag \\
& \qquad -\frac{1}{4S_i} \left( \frac{Q^i_{lk}-S_i (kl|kl)}{S_l} \right) W_{lk} \notag \\
&  = \frac{1}{4S_i} \frac{Q^i_{kl}}{S_k S_l} \tilde \tau_{kl}  + C'_{kl}
}
where $\tilde \tau_{kl} = W_{kl} S_l + W_{lk} S_k$ is the unnormalized `edge traffic' \cite{Aslyamov23}, and again $C'_{kl}$ is a quantity independent of $i$ that will drop out of 
\eq{
\frac{\p \log \pi_i}{\p F_{kl}} & = \left( \frac{Q^i_{kl}}{4 S_i S_k S_l} - \sum_q \frac{Q^q_{kl}}{4 S_q S_k S_l} \right) \tau_{kl} 
}
with $\tau_{kl}= \tilde \tau_{kl}/S$. \\

{\bf 3. Nonequilibrium Ensembles. }

First note that the quadratic form involving rates can be written
\eq{
\sum_{p,q} R_{pq} \ot_p \theta_q & = \sum_q \left[ \sum_{p, p\neq q} W_{pq} \ot_p \theta_q - \sum_{p, p \neq q} W_{pq} \ot_q \theta_q \right] \\
& = \sum_q \sum_{p, p\neq q} W_{pq} (\ot_p-\ot_q) \theta_q \\
& = \sum_q \sum_{p, p< q} W_{pq} (\ot_p-\ot_q) \theta_q + \sum_q \sum_{p, p> q} W_{pq} (\ot_p-\ot_q) \theta_q \\
& = \sum_{q,p, p< q} (\ot_p-\ot_q) (W_{pq} \theta_q - W_{qp} \theta_p) 
}

%{\red Check: suppose that $W_{pq} = w e^{E_q}$. Then this becomes
%\eq{
%\sum_{p,q} R_{pq} \ot_p \theta_q & = w \sum_{q,p, p< q} (\ot_p-\ot_q) (e^{E_q} \theta_q - e^{E_p} \theta_p) 
%}
%and under a change of variable $e^{E_q} \theta_q = \theta'_q$ we get 
%\eq{
%\pi_j = \frac{[\ot_j \theta_j]_{-R}}{\sum_k [\ot_k \theta_k]_{-R}} = e^{-E_j} \frac{J [\ot_j \theta'_j]_{-R'}}{J \sum_k e^{-E_k} [\ot_k \theta'_k]_{-R'}} = e^{-E_j}/Z' ,
%}
%where $R'$ has the same form as $R$ with all rates set to $w$, and $Z'$ here is independent of $j$, because $R'$ treats all states equally. Here $J=e^{\sum_k E_j}$ is the Jacobian from the change of variable.
%
%suppose that $W_{pq} = w e^{E_q-B_{pq}}$ where $B_{pq}=B_{qp}$. Then this becomes
%\eq{
%\sum_{p,q} R_{pq} \ot_p \theta_q & = w \sum_{q,p, p< q} (\ot_p-\ot_q) e^{B_{pq}} (e^{E_q} \theta_q - e^{E_p} \theta_p) 
%}
%
%
%}

In the SUSY approach a general observable of rate pairs can be written
\eq{
\ffrac{1}{N} \pi_ i \OO(W_{ij},W_{ji}) = \int \D\theta\D\ot \int \D s\D u \; e^{-R : (\ot \theta + i s u)} e^{t \ot_i \theta_i + i t s \cdot \1 \cdot u} \OO(W_{ij},W_{ji}) 
}
Our goal is to write this as the amplitude of a Feynman diagram \cite{Bachas94,DeGiuli19a}. States will be vertices of the diagram. First disentangle
\eq{
e^{i t s \cdot \1 \cdot u} & = e^{i t (\sum_k s_k) (\sum_k u_k)} \notag \\
& = \int \frac{dxdy}{\pi} e^{-(x^2+y^2)} e^{i \sqrt{t} (x+iy) \sum_k s_k + \sqrt{t} (x-iy) \sum_k u_k } \notag \\
& \equiv \int \frac{dxdy}{\pi} e^{-(x^2+y^2)} \prod_k \lambda(s_k,u_k) 
}
Since each state has an associated $\ot,\theta,s,u$ we are seeking a theory with 2 real coordinates and 2 Grassmann coordinates. Write $R = (\ot,\theta,s,u), T=(\oe,\eta,v,w)$ and $dR = ds du d\theta d\ot$. We have a $K$-component superfield
\eq{
\Phi^\mu(\ot,\theta,s,u) = \phi^\mu(s,u) + \ot \epsilon^\mu(s,u) + \bar\epsilon^\mu(s,u) \theta + \ot \theta \psi^\mu(s,u)
}
%\eq{
%\Phi(\ot,\theta,s,u) = \phi(s,u) + \ot \epsilon(s,u) + \bar\epsilon(s,u) \theta + \ot \theta \psi(s,u)
%}
where $\phi$ and $\psi$ are commuting and $\epsilon$ and $\bar\epsilon$ are anti-commuting. 

We seek a field theory whose diagrams give vertices a weight 
\eq{
\lambda(s,u) = \lambda(R) = e^{i \sqrt{t} (x+iy) s + \sqrt{t} (x-iy) u } .
}
and whose edges give the contributions in $e^{-R : (\ot \theta + i s u)}$ from each pair, i.e
\eq{
\zeta_{\mu\nu}(R,T; W_{\mu\nu},\oW_{\mu\nu}) = e^{-(\ot-\oe)(W_{\mu\nu}\eta-\oW_{\mu\nu} \theta)} e^{-i(s-v)(W_{\mu\nu} w-\oW_{\mu\nu} u)} .
}
for the transition from a $\nu$ state to a $\mu$ state. It is convenient to split these into intra-type $\mu=\nu$ transitions, and inter-type transitions $\mu\neq\nu$. The former can easily be obtained from a propagator $\zeta_{\mu}(R,T; W_{\mu\mu},\bar W_{\mu\mu})$ while the latter will be obtained from adding additional `activated' vertices.

%\eq{
%\zeta_{\mu}(R,T; W_{\mu\mu},\bar W_{\mu\mu}) = e^{-(\ot-\oe)(W_{\mu\mu}\eta-\oW_{\mu\mu} \theta)} e^{-i(s-v)(W_{\mu\mu} w-\oW_{\mu\mu} u)} .
%}
%\eq{
%\zeta(R,T; W,\bar W) = e^{-(\ot-\oe)(W\eta-\oW \theta)} e^{-i(s-v)(W w-\oW u)} .
%}

The special sites $i$ and $j$ will have their own special vertex. 

Since the propagator is not symmetric if any $\oW_{\mu\nu} \neq W_{\mu\nu}$, we need to use complex fields, which gives directed edges in its diagrams. Then $\Phi^\mu = \Phi_r^\mu + i \Phi^\mu_i$ where the real and imaginary parts are independent integration variables. 

Consider the action
%S(\Phi) & = \int dR \int dT \left[ \Phi(T) M(T,R)\Phid(R) - e^{t\ot \theta} \sum_{m,n \geq 1} \frac{\xi_{mn}}{m!n!} Re[\Phi(R)]^m Re[\Phi(T)]^n \lambda(R)\lambda(T) \zeta(R,T; W,\oW) \OO(W,\oW) \right] \notag\\
\eq{
S(\Phi) & = \sum_\mu \int dR \int dT \Phid^\mu(T) M_{\mu\mu}(T,R)\Phi^\mu(R) - \sum_\mu \int dR \sum_{n \geq 1} \frac{1}{n!} \xi_{\mu,n} \lambda(R)  [c_\mu\Phi^\mu(R) + c^\dagger_\mu \Phid^\mu(R)]^n \notag\\
&\qquad - \sum_{\mu \neq \nu} \xi_{\mu\nu} \int dR \int dT \lambda(R)\lambda(T) \Phid^\mu(T) \Phi^\nu(R) \zeta_{\mu\nu}(T,R)
}
and generating function
\eq{
\mathcal{W}(x,y,t, J; \OO) = \frac{1}{{\mathcal N}} \prod_{\mu \neq \nu} \oint' \frac{d\xi_{\mu\nu}}{\xi_{\mu\nu}} \prod_{\mu,n} \oint' \frac{d\xi_{\mu,n}}{\xi_{\mu,n}} \int \D\Phi e^{-S_{ext}(J)}
} 
where
\eq{
S_{ext}(J) = S + \sum_\mu N_\mu \sum_n \rho_{\mu,n} \log \xi_{\mu,n} + \sum_{\mu\neq \nu} E_{\mu\nu} \log \xi_{\mu\nu} - J \tilde \OO[\Phi]
}
is the extended action, and
\eq{
\tilde \OO[\Phi] & = \int dR \int dT e^{t\ot \theta} [c_{\mu_*}\Phi^{\mu_*}(R) + c^\dagger_{\mu_*} \Phid^{\mu_*}(R)]^{m_*} [c_{\nu_*}\Phi^{\nu_*}(T) + c^\dagger_{\nu_*} \Phid^{\nu_*}(T)]^{n_*} \lambda(R)\lambda(T) \notag\\
&\qquad \zeta_{\mu_*\nu_*}(R,T; W_{\mu_*\nu_*},\oW_{\mu_*\nu_*}) \OO(W_{\mu_*\nu_*},\oW_{\mu_*\nu_*}) .
}
Above $\sum_n \rho_{\mu,n} = 1$ for each $\mu$ and each $N_\mu\rho_{\mu,n}$ is a nonnegative integer. We absorb factors $2\pi i$ into $\oint' = \oint/(2\pi i)$. We explain $\mathcal{N}$ later.

Initially consider all $W_{\mu\nu} ,\oW_{\mu\nu} $ as fixed parameters. 

The Wick expansion with respect to the $\{\xi_{\mu,n}, \xi_{\mu\nu}\}$ generates diagrams as follows: each factor of $\xi_{\mu,n}$ gives a vertex of degree $n$ and type $\mu$, with an amplitude $\lambda$; each factor of $\xi_{\mu\nu}$ gives an activated vertex of type $\mu\nu$, with an amplitude $\lambda(R)\lambda(T) \zeta_{\mu\nu}(T,R)$; each factor of $J$ gives the special vertex pair with degrees $m_*$ and $n_*$ and an amplitude $e^{t\ot \theta}\lambda(R)\lambda(T) \zeta_{\mu_*\nu_*}(R,T; W_{\mu_*\nu_*},\oW_{\mu_*\nu_*}) \OO(W_{\mu_*\nu_*},\oW_{\mu_*\nu_*}) $; and each edge connecting states of the same type gets an amplitude
\eq{
\langle \Phi^\mu(T) \Phid^\nu(R) \rangle = \delta_{\mu\nu} M^{-1}_{\mu\mu}(T,R)
}
We assume that $\D \Phi$ includes a factor of the inverse square root of the functional determinant of $M$.
%\eq{
%\langle \Phi(T) \Phid(R) \rangle = M^{-1}(T,R)
%}
%Each factor of $J$ generates the special vertex for the chosen observable.

The coefficients $c_\mu,c^\dagger_\mu$ control the relative probability of $\Phi$ and $\Phid$ vertices. %By using $\Phi_r$ in the vertices we allow edges to be of $\Phi$ or $\Phid$ type with equal probability (this can be generalized). 
The numbers $\rho_{\mu,n}$ and the contour integrals specify the degree distribution of the state space. There are $N+2$ states in total, including the two from the special vertex, but excluding activated states.

 We want $M^{-1}_{\mu\mu}(T,R) = \zeta_{\mu\mu}(T,R)$, i.e. we want
%\eq{ \label{G1}
%\delta(R-T) \delta_{\mu\nu}&= \delta(s-v)\delta(u-w)(\ot-\oe)(\theta-\eta) \delta_{\mu\nu} \\
%& = \sum_\kappa \int dT' \; M_{\mu\kappa}(R,T') \zeta_{\kappa\nu}(T',T)
%}
% We want $M^{-1}(T,R) = \zeta(T,R)$, i.e. we want
\eq{ \label{G1}
\delta(R-T) &= \delta(s-v)\delta(u-w)(\ot-\oe)(\theta-\eta) \notag \\
& = \int dT' \; M_{\mu\mu}(R,T') \zeta_{\mu\mu}(T',T)
}
If we can find such an $M$ then the above generating function measures the observable $\OO$ over an ensemble of random state spaces. %With contour integrals we can specify the number of each type of vertex, and thus the entire degree distribution. In particular from the $\xi_{mn}$ we probably just want one. 

There are $N_\mu$ states of type $\mu$, with $E'_\mu= \sum_{\nu \neq \mu} [E_{\mu\nu}+E_{\nu\mu}]$ transitions to states of other types, and 
%have $N_\mu \sum_n \rho_{\mu,n} + \sum_{\nu \neq \mu} [M_{\mu\nu}+M_{\nu\mu}] = 2 M''_\mu$ where $M''_\mu$ is the number of transitions to states of the same type (including self-transitions).
 $E''_\mu = \half \sum_{\nu \neq \mu} [E_{\mu\nu}+E_{\nu\mu}] + \half N_\mu \sum_n \rho_{\mu,n} n$ transitions to other states of the same type (including self-transitions).

Note that $\mathcal{W}$ generates both connected and disconnected diagrams, but Kirchoff's theorem requires (A2 in the main text) that the state space be connected; moreover we want a connected component with a single special vertex. We notice that if a connected component lacks a special vertex, then it will give a vanishing contribution, since its fermionic part will involve a factor $\det(-R) = 0$ (lacking the $e^{t\ot_i \theta_i}$ factor that makes it finite). In particular $\mathcal{W}(J=0) = 0$. We want the part linear in $J$, i.e. $\lim_{J \to 0} \mathcal{W}(J)/J = \lim_{J \to 0} \p \mathcal{W}/\p J$ \footnote{The usual argument relating $\mathcal{W}$ to $\log \mathcal{W}$ breaks down because $\lim_{J\to 0} \log \mathcal{W}$ diverges.}.  

Finally, the generating function overcounts the observable by the number of diagrams $\mathcal{N}$ in the ensemble (with the $c,c^\dagger$ weights). This can be written as a similar but simpler integral
\eq{
\mathcal{N} = \prod_{\mu \neq \nu} \oint' \frac{d\xi_{\mu\nu}}{\xi_{\mu\nu}^{1+E_{\mu\nu}+\delta_{\mu,\mu_*}+\delta_{\nu,\nu_*}}} \prod_{\mu,n} \oint' \frac{d\xi_{\mu,n}}{\xi_{\mu,n}^{1+N_\mu\sum_n \rho_{\mu,n}}} \int D\phid D\phi e^{-\sum_\mu \phid^\mu \phi^\mu + \sum_\mu \sum_{n\geq 1} \xi_{\mu,n} [c_\mu \phi^\mu + c_\mu^\dagger \phid^\mu]^n /n! + \sum_{\mu\neq\nu} \xi_{\mu\nu} \phid^\mu\phi^\nu}
} 
where here $\phi$ is simply a $K-$component complex vector. We are assuming that the $\{ E,N,\rho \}$ are such that the fraction of disconnected diagrams is negligible in the $N \to \infty$ limit.
%second, the Gaussian integrations will include a factor that is the square root of the functional determinant of the propagator, $(\det M)^{1/2}$. 

%thus we need to take the logarithm of the generating function to extract the desired diagrams; this requires that we normalize the measure over $\Phi$ such that 
%\eqs{
%\int \D\Phi e^{-S_{ext}|_{\xi_{\mu,n} \to 0,\xi_{\mu\nu} \to 0}} = 1
%}
%Then $\log\mathcal{W}$ generates connected diagrams. At order $J$ these will have one insertion of the observable, as desired.

Putting everything together we have
\eq{
\ffrac{1}{N} \langle \pi_{\mu_*} \OO(W_{\mu_*\nu_*},\oW_{\mu_*\nu_*})  \rangle = \int \frac{dxdy}{\pi} e^{-x^2-y^2} \left.\frac{\p \mathcal{W}(x,y,t, J; \OO)}{\p J}\right|_{J=0}
}
where $\pi_{\mu_*}$ is the probability to be in a particular state of type $\mu_*$. Note that as $N \to \infty$ at finite $K$, then $\mathcal{W}$ will be dominated by a saddle-point\footnote{If $K\to\infty$ then a more delicate analysis is necessary. For example if $K\sim N^{1/2}$ and each $N_\mu \sim N^{1/2}$ then $N=\sum_\mu N_\mu \sim N^{2/2}$ is consistent. }. Thus 
\eq{
\mathcal{W}(x,y,t,J; \OO) \sim \frac{1}{\mathcal{N}}\sum_{SP} \ffrac{1}{\prod_{\mu,n}(2\pi i\xi_{\mu,n})\prod_{\mu,\nu}(2\pi i\xi_{\mu\nu})} e^{-S_{ext}(J)} (S_{ext}(J)'')^{-1/2}
}
where the sum is over all saddle points, all quantities are evaluated on the saddle-point, and we use $S_{ext}''$ to schematically denote the functional determinant of the Hessian. We include in this factor the Pfaffian that is obtained by integrating out the fermionic fields $\epsilon,\bar\epsilon$. The factor of $J$ needed to make $W/J$ finite as $J \to 0$ comes from the Hessian; in the action and in the saddle-point equations, $J$ can be set to 0. Thus all rate-pair observables are determined by the same equations, as expected.

Under some conditions this can be simplified considerably. From normalization we have
\eq{
1 = \sum_\mu N_\mu \langle \pi_\mu \rangle =  N \sum_\mu N_\mu \int \frac{dxdy}{\pi} e^{-x^2-y^2} \left.\frac{\p \mathcal{W}(x,y,t, J; 1)}{\p J}\right|_{J=0} .
}
Comparing with the original expression for $\mathcal{W}$, we see that the derivative $\p {\mathcal W}/\p J$ pulls down $\tilde \OO$. For normalization the latter is
\eq{
\tilde \OO_{1,\mu}[\Phi] = \int dR \int dT e^{t\ot \theta} [c_{\mu}\Phi^{\mu}(R) + c^\dagger_{\mu} \Phid^{\mu}(R)]^{m_*} [c_{\nu_*}\Phi^{\nu_*}(T) + c^\dagger_{\nu_*} \Phid^{\nu_*}(T)]^{n_*} \zeta_{\mu\nu_*}(R,T; W_{\mu\nu_*},\oW_{\mu\nu_*}) 
}

If a single saddle-point dominates (the usual case) and if $\mathcal{W}$ is independent of $x,y$, which means that SUSY is preserved, then the prefactors can all be cancelled and we have simply
\eq{
\langle \pi_{\mu_*} \OO(W_{\mu_*\nu_*},\oW_{\mu_*\nu_*})  \rangle = \frac{\tilde \OO[\Phi]}{\sum_\mu N_\mu \tilde \OO_{1,\mu}[\Phi]  } \qquad \text{SUSY preserved} 
}
where all quantities are evaluated on the saddle point.

%Moreover, if a single saddle-point dominates, then for any pair of observables, schematically 1 and 2, we have 

%Then we obtain the key result
%\eq{
%\ffrac{1}{N} \langle \pi_{\mu_*} \OO(W_{\mu_*\nu_*},\oW_{\mu_*\nu_*})  \rangle \sim \int \frac{dxdy}{\pi } e^{-x^2-y^2} \tilde \OO[\Phi] 
%}
%This is valid for any observable of the given form.

%Note that in the special case $\OO = 1$ we have
%\eq{
%1/N^2 = \int \frac{dxdy}{\pi i} e^{-(x^2+y^2)} \mathcal{W}(x,y,t; 1) 
%}
%Therefore if $S_{ext} \neq 0$ and there is a single dominant saddle point then we obtain the key result
%\eq{
%\langle \pi_ i \OO(W_{ij},W_{ji}) \rangle \sim \ffrac{1}{N} \tilde\OO[\Phi]
%}
%where the right-hand side is precise asymptotic as $N\to\infty$, evaluated on the saddle point. 

To see why saddle-point is justified very generally, note that if, say, all $N_\mu \sim N \gg 1$ then we can freely rescale all $\xi_{\mu\nu},\xi_{\mu,n}$ without changing the scaling of the $\log \xi_{\mu\nu}, \log \xi_{\mu,n}$ terms in $S_{ext}$. This freedom allows us to ensure that the terms in $S$ with these variables scale as $N$, and the remaining freedom in scaling $\Phi$ ensures that the kinetic term also scales as $N$. Letting $\xi_{\mu\nu} \sim N^{a_{\mu\nu}}, \xi_{\mu,n} \sim N^{a_{\mu,n}}, \Phi \sim N^a$ we get $1 = a_{\mu\nu}+2a = a_{\mu,n} + na = 2a$ which are trivially solved.

Moreover, we can also allow the rates $W_{\mu\nu},\oW_{\mu\nu}$ to fluctuate: as long as rate pairs are independent, in \eqref{G1} and in the activated vertices we simply replace $\zeta$ by $\langle \zeta \rangle$ over the rate ensemble, and in the special vertex we replace $\zeta(R,T; W_{\mu\nu},\oW_{\mu\nu}) \OO(W_{\mu\nu},\oW_{\mu\nu})$ by $\langle \zeta(R,T; W_{\mu\nu},\oW_{\mu\nu}) \OO(W_{\mu\nu},\oW_{\mu\nu}) \rangle$. To allow the $\mu-\mu$ transitions to have rate fluctuations, the simplest approach is to find the inverse propagator $M$ for constant rates, and then add rate fluctuations through a $\xi_{\mu\mu}$ term, with random rates, say $W'_{\mu\mu},\oW'_{\mu\mu}$. These random rates will renormalize the effective $\mu-\mu$ rates through a Dyson equation $M_{eff}(R,T) = M(R,T) - \xi_{\mu\mu} \lambda(R)\lambda(T) \langle \zeta'_{\mu\mu}(R,T)\rangle$.

%Finding the inverse propagator $M$ over a general ensemble may be difficult (note that $M$ cannot depend on random rates).

Now we solve the propagator equation \eqref{G1}, where we suppress the $\mu$ indices. Taking a Fourier transform with $Q = (p_1,p_2,q_1,q_2)$, where $q_j$ are bosonic and $p_j$ are fermionic, we have
\eqs{ %\label{G2}
e^{i R \cdot Q} & = \int dT e^{iT\cdot Q} \delta(R-T) \notag\\
&= \int dv dw \int d\eta d\oe \; e^{ivq_1 + i w q_2 + i \oe p_1 + i \eta p_2 } \delta(s-v)\delta(u-w)(\ot-\oe)(\theta-\eta) \notag\\
& = \int dv dw \int d\eta d\oe \; e^{ivq_1 + i w q_2 + i \oe p_1 + i \eta p_2} \int dT' \; M(R,T') \langle e^{-(\oe'-\oe)(W\eta-\oW \eta')} e^{-i(v'-v)(W w-\oW w')} \rangle \\
& = \int dv dw \int d\eta \; e^{ivq_1 + i w q_2 + i \eta p_2} \int dT' \; M(R,T') \langle e^{-\oe'(W\eta-\oW \eta')} e^{-i(v'-v)(W w-\oW w')} (ip_1 + W \eta - \oW \eta') \rangle \\
& = \int dv dw \; e^{ivq_1 + i w q_2} \int dT' \; M(R,T') \langle e^{\oe' \oW \eta')} e^{-i(v'-v)(W w-\oW w')} [ (ip_2+W \oe') (ip_1 - \oW \eta') + W ]\rangle \\
& = 2\pi \int dv \; e^{ivq_1} \int dT' \; M(R,T') \langle e^{\oe' \oW \eta')} e^{i(v'-v)\oW w'} [ (ip_2+W \oe') (ip_1 - \oW \eta') + W ] \delta(q_2-(v'-v)W) \rangle \\
& = 2\pi \int dT' \; M(R,T') \langle \ffrac{1}{W} e^{i(v'-q_2/W) q_1}  e^{\oe' \oW \eta')} e^{i q_2 \oW w'/W} [ (ip_2+W \oe') (ip_1 - \oW \eta') + W ] \rangle \\
& = 2\pi \int dT' \; M(R,T') \langle e^{i(v'-q_2/W) q_1}  e^{\oe' \oW \eta')} e^{i q_2 \oW w'/W} e^{(ip_2+W \oe') (ip_1 - \oW \eta')/W} \rangle \\
& = 2\pi \int dT' \; M(R,T') \langle e^{iv'q_1-i q_2 q_1/W}  e^{i q_2 w' \oW /W} e^{-p_2 p_1/W + \oe' ip_1  - ip_2 \eta' \oW /W} \rangle
}
Write
\eqs{
M(R,T') & = M_1 (1  + \oe' M_2 + M_3 \eta' + \oe' \eta' M_4 ) \\
& = M_1 e^{\oe' M_2 + M_3 \eta' + \oe' \eta' (M_4 + M_2 M_3) }
}
giving
\eq{ \label{G3}
e^{i R \cdot Q} & = 2\pi  \int dv'dw' \; M_1 \int d\eta'd\oe' \langle e^{\oe' (M_2+ip_1) + (M_3-ip_2\oW/W) \eta' + \oe' \eta' (M_4 + M_2 M_3) } e^{iv'q_1 +i q_2 w' \oW /W -i q_2 q_1/W -p_2 p_1/W} \rangle \notag\\
& =  2\pi \int dv'dw' \; M_1 (M_4 + M_2 M_3) \langle e^{- (M_3-ip_2\oW/W) (M_2+ip_1)/(M_4+M_2M_3) } e^{iv'q_1 +i q_2 w' \oW /W -i q_2 q_1/W -p_2 p_1/W}  \rangle 
}
where we assumed that $M_4+M_2M_3 \neq 0$, but if this vanishes then one can take the limit from this expression.

So far we could allow $W_{\mu\mu}$ and $\oW_{\mu\mu}$ to be random, but as explained above, it is easier to obtain such disorder through a vertex. Moreover even if such a vertex is neglected, there is still topological disorder, and we can still allow arbitrary disorder in the $\mu\neq\nu$ transitions. 

So we let $W_{\mu\mu}$ and $\oW_{\mu\mu}$ be constant. The fermionic and bosonic parts then decouple. The former requires
\eqs{
e^{i \ot p_1 + i \theta p_2} \propto (M_4 + M_2 M_3) e^{- (M_3-ip_2\oW/W) (M_2+ip_1)/(M_4+M_2M_3) } e^{-p_2 p_1/W} 
}
which implies $\oW/(M_4+M_2M_3)+1=0$ and then
\eqs{
e^{i \ot p_1 + i \theta p_2} \propto -\oW e^{M_3 M_2/ \oW} e^{M_3 ip_1 /\oW }  e^{-ip_2 M_2/W} 
}
so that $M_2 = W\theta$ and $M_3 = \ot \oW$. Then \eqref{G3} becomes
\eqs{
e^{isq_1 + iu q_2} & =  -2\pi \oW e^{W \ot \theta} \int dv'dw' \; M_1 e^{iq_1v' + i q_2 w'\oW/W} e^{-iq_1 q_2/W} 
}
Consider $M_1 = (2\pi)^{-1} \delta(s-v')\delta(b u-w') m e^{c \p_{v'} \p_{w'}}$. This becomes
\eqs{
e^{isq_1 + iu q_2} & = -\oW e^{W \ot \theta} \int dv'dw' \; \delta(s-v')\delta(b u-w') m e^{c \p_{v'} \p_{w'}} e^{iq_1v' + i q_2 w'\oW/W} e^{-iq_1 q_2/W} \\
&= -\oW e^{W \ot \theta} \int dv'dw' \; \delta(s-v')\delta(b u-w') m e^{c (iq_1) (iq_2 \oW/W)} e^{iq_1v' + i q_2 w'\oW/W} e^{-iq_1 q_2/W} \\
&= -\oW e^{W \ot \theta} m e^{-c q_1 q_2 \oW/W} e^{iq_1 s + i q_2 b u \oW/W} e^{-iq_1 q_2/W}
}
which works provided $c = -i/\oW, b = W/\oW, m = -e^{-W \ot \theta}/\oW$. The final result is
%\eq{
%M(R,T') = -\frac{2\pi }{\oW} \delta(v'-s)\delta(w'- u W/\oW) e^{-(\ot-\oe') (W\theta-\oW \eta')} e^{-i \p_{v'} \p_{w'} / \oW}
%}
\eq{
M_{\mu\mu}(R,T') = -\frac{2\pi }{\oW_{\mu\mu}} \delta(v'-s)\delta(w'- u W_{\mu\mu}/\oW_{\mu\mu}) e^{-(\ot-\oe') (W_{\mu\mu}\theta-\oW_{\mu\mu} \eta')} e^{-i \p_{v'} \p_{w'} / \oW_{\mu\mu}} .
}

Finally, let us show how SUSY manifests in the field theory. 
\newcommand{\Qd}{Q^{\dagger}}
The SUSY generators are
\eq{
Q & = \theta \cdot \nabla_u - i s \cdot \nabla_{\ot} + \eta \cdot \nabla_w - i v \cdot \nabla_{\oe} \\
\Qd & = \ot \cdot \nabla_s + i u \cdot \nabla_{\theta} + \oe \cdot \nabla_v + i w \cdot \nabla_{\eta}
}
Now note that under the change of variable ($\chi$ is a Grassmann scalar)
\eqs{
\ot & \to \ot + i s \chi \\ 
u & \to u + \theta \chi
}
a function $f(\ot,u)$ becomes $f + i s \chi \p_{\ot} f + \theta\chi \p_u f = f - \chi Q f$. It follows that if we generate a variation in $\Phi,\Phid$ by $Q$ in the action, all dependence via these variables can be soaked up by a change of variable in the integral. Indeed the variation $\delta \Phi = \chi Q \Phi, \delta \Phid = \chi Q \Phid$ modifies an integrand $g(\Phi,\Phid)$ to $g(\Phi,\Phid) + \p_\Phi g \delta \Phi + \p_\Phid g \delta \Phid$ and the new terms are cancelled by those from a change of variable. However, under this change of variable, new terms will be introduced through any additional dependence on $R,T$. So let us check that the kernel $\zeta$ is invariant:
\eqs{
Q \zeta_{\mu\nu} = \theta i (s-v) \oW_{\mu\nu} \zeta_{\mu\nu} + is (W_{\mu\nu}\eta-\oW_{\mu\nu} \theta) - \eta i (s-v) W_{\mu\nu} -iv (W_{\mu\nu}\eta-\oW_{\mu\nu} \theta) = 0
}
and the $\delta$-function 
\eqs{
Q \delta(s-v)\delta(u-w)(\ot-\oe)(\theta-\eta) = \delta(s-v) (\theta-\eta) \delta'(u-w)(\ot-\oe)(\theta-\eta) - \delta(s-v)\delta(u-w) (is-iv) (\theta-\eta) = 0
}
so that $M$ is also invariant. As a result, only the dependence via $\lambda(R)$ remains; this term breaks SUSY as expected. 

%\eq{
%\delta \Phi(\ot,\theta,s,u) & = \chi Q \Phi = \chi \left[ \theta \nabla_u (\phi + \ot \epsilon) - i s (\epsilon + \theta \psi) \right] \\
%\delta \Phid(\ot,\theta,s,u) & = \chi Q \Phid = 
%}

\end{widetext}

\end{document}